\documentclass[12pt,preprintnumbers,amsmath,amssymb]{revtex4-1}
\pdfoutput=1
\usepackage{graphicx}
\usepackage{color}
\usepackage[bookmarksnumbered=true,colorlinks=true,linkcolor=black,citecolor=black]{hyperref}

\newcommand{\beq}{\begin{equation}}
\newcommand{\eeq}{\end{equation}}
\newcommand{\bseq}{\addtocounter{subeqno}{1}\begin{subequations}}
\newcommand{\eseq}{\end{subequations}}

\font\mathscript=eusm10 at 12pt
\font\mathscripts=eusm7
\font\mathscriptss=eusm5
\newfam\mathscri
\textfont\mathscri=\mathscript
\scriptfont\mathscri=\mathscripts
\scriptscriptfont\mathscri=\mathscriptss
\def\mathscr#1{{\fam\mathscri\relax#1}}

\font\mathfrakt=eufm10 at 12pt
\font\mathfrakts=eufm7
\font\mathfraktss=eufm5
\newfam\mathfraki
\textfont\mathfraki=\mathfrakt
\scriptfont\mathfraki=\mathfrakts
\scriptscriptfont\mathfraki=\mathfraktss
\def\mathfrak#1{{\fam\mathfraki\relax#1}}

\def\CO{{\cal O}}

\def\half{{\textstyle{1\over 2}}}
\def\pa{\partial}

\newcommand{\e}{{\rm e}}

\def\ln{{\rm ln}}

\begin{document}

\title{Davies Critical Point and Tunneling}
\author{HoSeong La}
\email[Present address: Department of Physics and Astronomy,
Vanderbilt University, Nashville, TN 37235, USA]{}
\affiliation{
Department of Physics and Astronomy, 
The University of Tennessee, 
Knoxville, TN 37996, USA\\
{\tt hsla.avt@gmail.com}
}

\begin{abstract}
From the point of view of tunneling, the physical meaning of 
the Davies critical point of a second order phase transition 
in the black hole thermodynamics is clarified. At the critical point,
the nonthermal contribution vanishes so that the black hole radiation
is entirely thermal. It separates two phases: one with radiation enhanced
by the nonthermal contribution, the other suppressed by 
the nonthermal contribution. 
We show this in both charged and rotating black holes.
The phase transition is also analyzed in the cases in which
emissions of charges and angular momenta are incorporated.
\end{abstract}


\maketitle

\setcounter{footnote}{0}
\setcounter{page}{1}
\setcounter{section}{0}
\setcounter{subsection}{0}
\setcounter{subsubsection}{0}


\section{Introduction}

Upon the discovery of the thermodynamic interpretation of 
black holes\cite{Bekenstein:1973ur}, 
the first critical phenomenon identified was
the second order phase transition discovered 
by Davies more than thirty years ago\cite{Davies:1978}. 
The phase structure was analyzed by Hut\cite{Hut},
and it has been generalized for varying charges 
of isolated charged black holes\cite{Hiscock:1990ex}.
The issue of stability in the Kerr black hole case
has been addressed\cite{Kaburaki:1993ah}. 
Yet, all these are still in the thermal radiation context based on the 
Hawking process.
Here we present a new proposal to clarify the physical meaning of 
this critical phenomenon in the context of the nonthermal radiation of 
Parikh-Wilczek\cite{Parikh:1999mf}\cite{Parikh:2004rh}. 
The existence of nonthermal radiation from black holes have been 
anticipated\cite{Damour:1974qv}\cite{Blandford:1977ds} 
beyond Hawking's thermal radiation\cite{Hawking:1974sw}.

Black holes are unusual thermodynamic systems because we cannot see
any structural changes directly as far as critical phenomena are concerned.
The entropy changes smoothly so that checking the behavior of the horizon
cannot tell us either.
So the only way to tell what is going on in the black hole phases is
if there is any difference in emissions of information from different phases.
So the tunneling argument of Parikh-Wilczek is a good candidate to 
investigate if there are such emissions telling about the black holes phases.
In this paper, we find that indeed this is the case, 
and that, using this information, we can explain
the physical meaning of the Davies critical phenomenon.

In short, we find that there is a competition between thermal part of radiation
and nonthermal part. At the critical point, the nonthermal part vanishes,
leaving only thermal radiation that peaks. 
The critical point separates two phases:
In one phase, the nonthermal contribution
enhances the total radiation, while in the other phase 
the nonthermal contribution actually suppresses the total radiation.
Once we extend to the cases of emissions of charges or angular momenta,
we can observe over all enhancement or suppression, but the peak of emission
remains at (or near) the critical point and separation of two phases persists.
We could check the characteristics of each phase of black holes with respect to
the radiation and emissions of charges and angular momenta
by introducing effective free energies. 

This paper is organized as following. In section 2, we review and identify
what the Davies critical point is. In section 3, we explain this critical phenomenon in the RN (Reissner-Nordtr\"om) black hole case, 
using the tunneling argument. Then, in section 4, it is extended to incorporate
the emissions of charges and angular momenta in the KN (Kerr-Newman) case.
Finally, in section 5, some further comments are given.

\section{Davies Critical Point}

The Davies critical point is identified by the singular behavior of 
specific heat at some nontrivial value of charge-to-mass ratio, $Q/M$, or, 
angular-momentum-to-mass ratio, $J/M^2$, away from the extremal 
limit\cite{Davies:1978}. A second order phase transition occurs at this 
critical point and the phenomenon is generic 
for any charged or rotating black holes. 
The nature of this critical phenomenon is not entirely clear except it has been
known that the specific heat changes the sign abruptly.

\begin{figure}[t]
\centering                                                        
\begin{minipage}[b]{0.45\textwidth}
\centering                                                        
\includegraphics[width=0.58\textwidth]{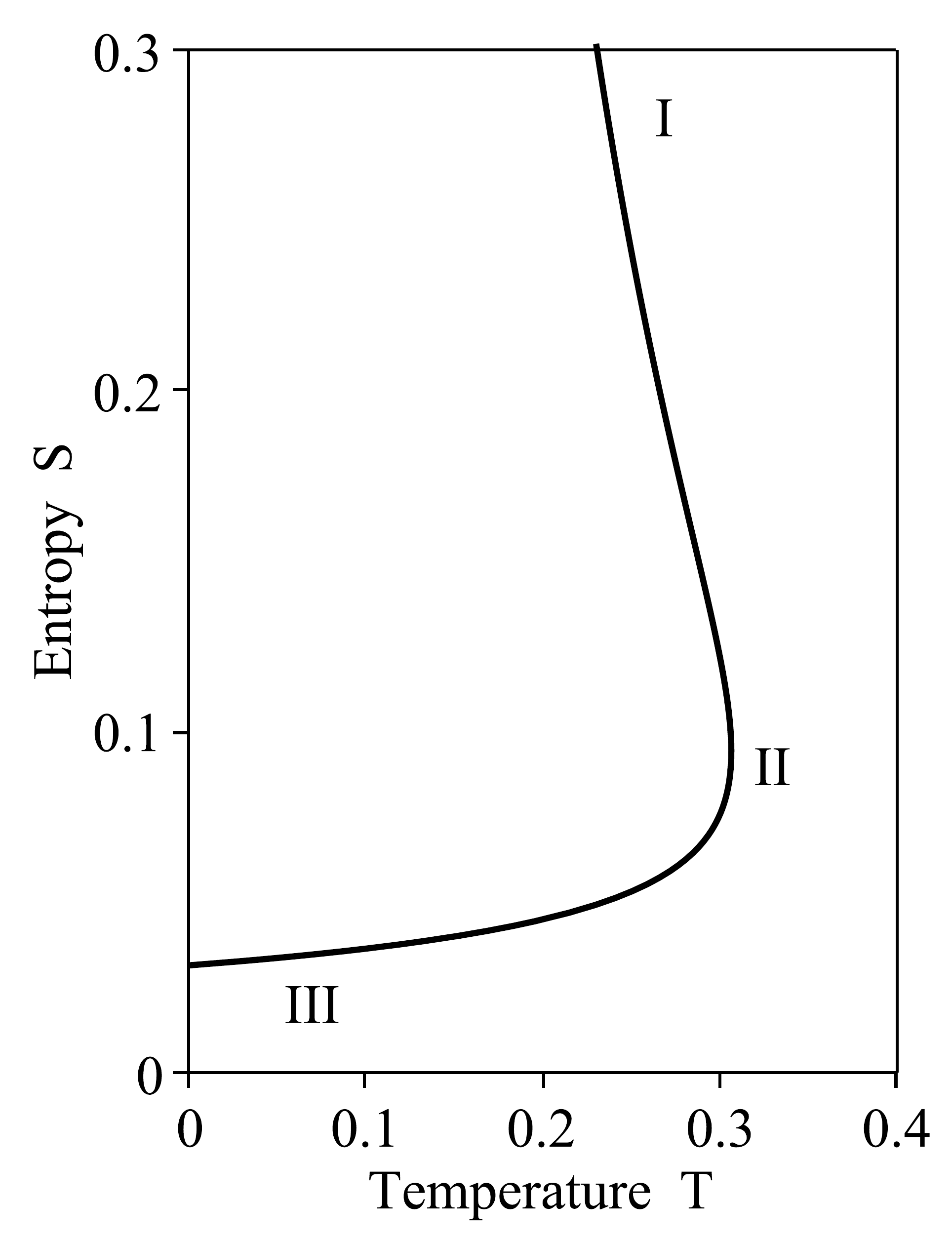}
\parbox{0.9\textwidth}{
\caption{S vs. T plot of RN black hole for fixed $Q=0.1$: 
   The turning point II is Davies critical point}
   \label{fig:rnst}
}
\end{minipage}%
\hspace{0.04\textwidth}%
\begin{minipage}[b]{0.45\textwidth}
\centering                                                         
\includegraphics[width=0.58\textwidth]{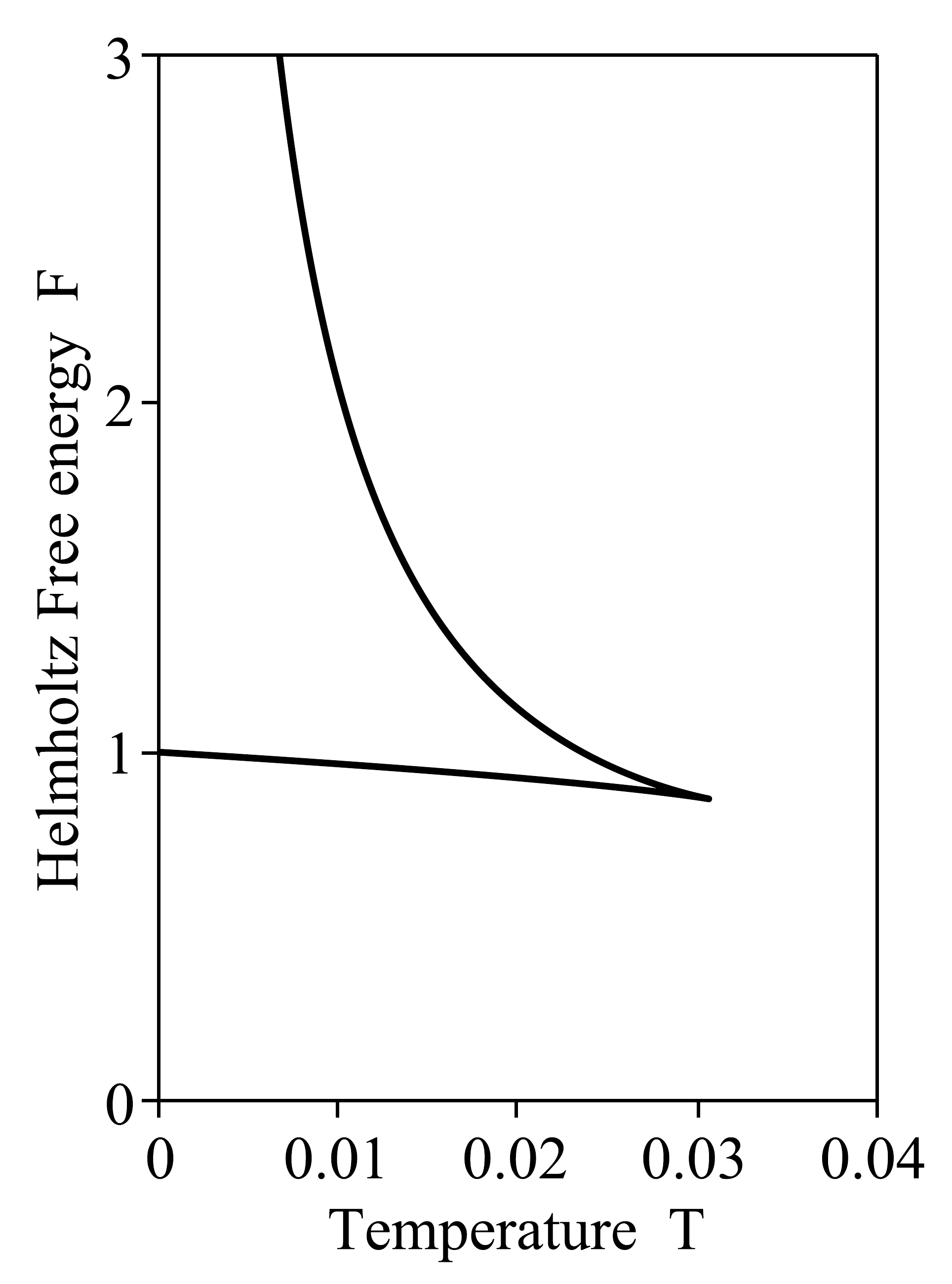}
\parbox{0.9\textwidth}{
   \caption{Helmholtz free energy of RN black hole for fixed $Q=1$: 
   The end point of the cusp is Davies critical point}
   \label{fig:rnhf}
}
\end{minipage}
\end{figure}

To illustrate the Davies critical point, we shall start with the RN 
black hole. Fig.\ref{fig:rnst} shows the relationship between
the Bekenstein-Hawking entropy 
and the Hawking temperature  for some value of charge $Q$, and,
in this case, the Davies critical point is the turning point marked by II.
At the critical point, $|Q|/M_c = \sqrt{3}/2$,
$T_c = 1/(9\pi M_c)$ and $S_c = (9/4)\pi M_c^2$ for a given charge $Q$.
As $Q$ varies, the critical points trace along $S_c\propto 1/T_c^2$.
As one can see in the plot, specific heat, which is related 
to the slope of the curve, $\pa S/\pa T$,
is singular at II, indicating the occurrence of some kind of a second order
phase transition. The specific heat is negative in region I and positive in 
region III.
The critical exponent for the specific heat is 1/2 such that, 
as $T_H$ approaches $T_c$,
\beq
|c_Q| \sim (T_c -T_H)^{-1/2}.
\eeq

\begin{figure}[t]
\centering                                                        
\begin{minipage}[b]{0.45\textwidth}
\centering                                                        
\includegraphics[width=0.58\textwidth]{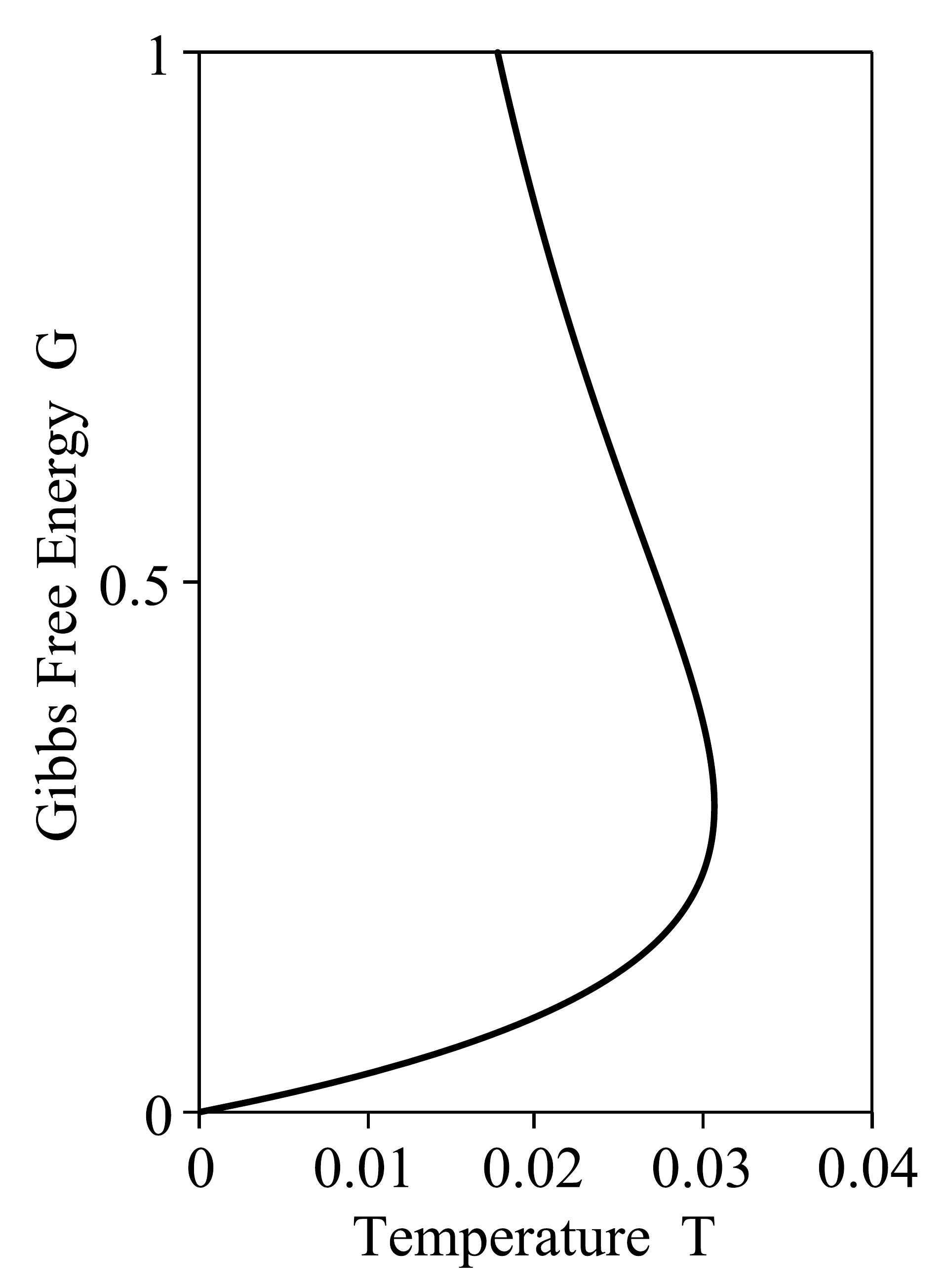}
\parbox{0.9\textwidth}{
\caption{RN case: At $T=0$, $G=0$.}
\label{fig:rn_gf}
}
\end{minipage}%
\hspace{0.04\textwidth}%
\begin{minipage}[b]{0.45\textwidth}
\centering                                                         
\includegraphics[width=0.58\textwidth]{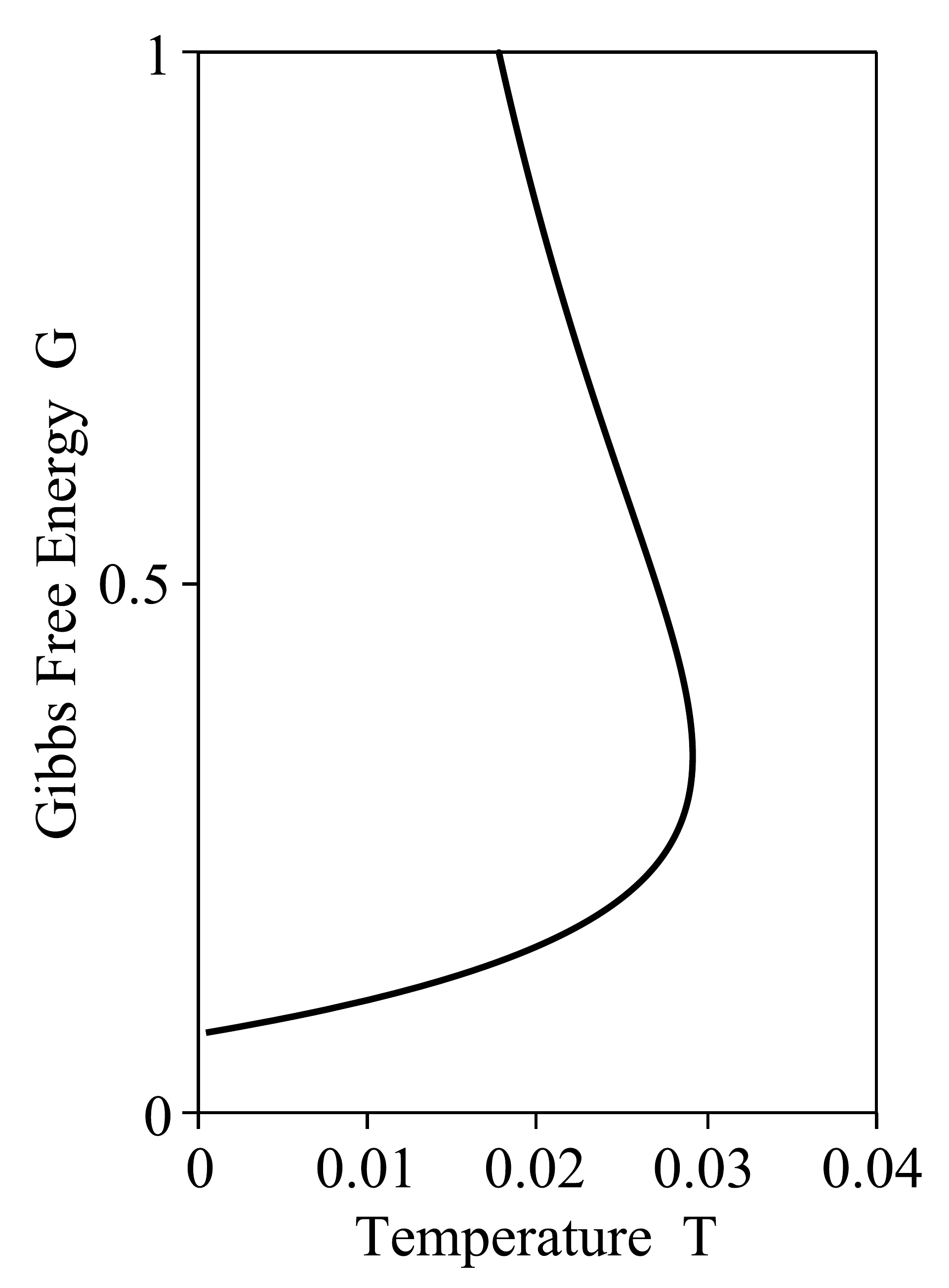}
\parbox{0.9\textwidth}{
\caption{KN case: At $T=0$, $G>0$.}
\label{fig:kn_gf}
}
\end{minipage}
\end{figure}

The Helmholtz free energy of RN black hole (fig.\ref{fig:rnhf})
shows a cusp formed at the Davies critical point. Note that,
unlike normal thermodynamic systems, the part with negative specific heat
of Helmholtz free energy is convex. The Gibbs free energy also displays
a turning point at the Davies critical point. 
In the extremal limit,
Gibbs free energy vanishes for RN black hole (fig.\ref{fig:rn_gf}), 
but in the KN case Gibbs free energy no longer vanishes 
due to the rotational degrees of freedom (fig.\ref{fig:kn_gf}).

Compared with the Schwarzschild black hole case, in which the specific heat
is always negative and temperature does not have a turning point,
in the RN or KN case there must be something happening to generate such a
turning point that the behavior of the temperature changes, at the same time,
changing the specific heat from negative to positive.

\section{Davies Critical Point from Tunneling 
in Reissner-Nordstr\"om Black Hole}

The appearance of such a charge-to-mass ratio as a critical point away 
from the extremal limit is quite intriguing and it certainly calls for 
deeper understanding of its physical implications. 
In this paper,
We argue that what happens is due to the nonthermal black hole radiation.
This can be seen by using Parikh-Wilczek's 
tunneling argument\cite{Parikh:1999mf} of black hole radiation as follows.

In tunneling, the emission rate is given by
\begin{align}
\label{e:n1}
\Gamma&\sim {\rm exp}\left(-2\,{\rm Im} I\right) \nonumber\\
&={\rm exp}\left(-2\,{\rm Im}\int p_r dr\right) \nonumber\\
&={\rm exp}
\left({-2\,{\rm Im}\int_M^{M-\omega}\half\beta_{\rm M}(H)dH}\right),
\end{align}
where 
\beq
\beta_{\rm M}(H)\equiv 2\int_{r_{\rm in}}^{r_{\rm out}}{dr\over \dot{r}}
\eeq 
is related to the inverse Hawking temperature as we shall see later.
After integrating, which is done by Parikh-Wilczek, we get
\beq
\label{e:n2}
\Gamma \sim \e^{-\beta_{\rm eff}(\omega)\omega},
\eeq
where
\begin{align}
{1\over 2\pi}\beta_{\rm eff}(\omega)  
\equiv -{1\over\omega}
\bigg(
&(M-\omega)^2  +(M-\omega)\sqrt{(M-\omega)^2-Q^2} \nonumber\\
&-M^2 -M\sqrt{M^2 -Q^2}
\bigg).
\end{align}
$T_{\rm eff} =1/\beta_{\rm eff}$ is called the effective temperature,
which is interpreted as the indication how temperature changes after
the emission of radiation. Note that this effective temperature is not
the temperature after the emission of radiation, but it is a quantity
still defined at the temperature before the emission.
Assuming $1\gg \omega/M$ and $1\gg (2\omega M-\omega^2)/(M^2 -Q^2)$,
we can expand $\beta_{\rm eff}(\omega)$ as
\beq
\label{e:I}
{1\over 2\pi}\beta_{\rm eff}(\omega) 
={1\over 2\pi T_H} - \omega f(M,Q) + {\rm h.o.},
\eeq
where 
\beq
T_H=1/\beta_H
={1\over 2\pi}{\sqrt{M^2-Q^2}\over \left(M+\sqrt{M^2-Q^2}\right)^2}
\eeq
is the Hawking temperature and
\beq
f(M,Q) \equiv 1 + {3\over 2}{M\over (M^2-Q^2)^{1/2}}
-{1\over 2}{M^3\over (M^2-Q^2)^{3/2}}.
\eeq
The key observation here is that the sign of $f(M,Q)$ depends on $M$ and $Q$.
In fact, $f(M,Q)=0$ for $|Q|/M_c=\sqrt{3}/2$, which is nothing but the Davies 
critical value! Note that $f(M,Q) \gtrless 0$ for $M \gtrless M_c$.

If we can relate $f(M,Q)$ to the specific heat, we should be able to 
understand this better. Indeed, in terms of specific heat 
\begin{align}
c_Q &\equiv {1\over M}\left({\pa M\over\pa T}\right)_Q \nonumber\\
&=2\pi {(M+\sqrt{M^2-Q^2})^2\sqrt{M^2-Q^2}\over M^2-2M\sqrt{M^2-Q^2}}\nonumber\\
&=-{\beta_H^2\over4\pi M} {1\over f(M,Q)},
\end{align}
we can reorganize eq.(\ref{e:I}) as
\beq
\label{eq:beff}
\beta_{\rm eff}(\omega) 
= \beta_H +{1\over 2}{\omega\over M}{\beta_H^2\over c_Q}+ \CO(\omega^2).
\eeq
This clearly shows how Davies critical point appears in the tunneling
context.
As we shall show later, this is generic for any black hole with 
the corresponding specific heat.
Now one can see that, for $c_Q<0$, $\beta_{\rm eff}<\beta_H$ so that
(effective) temperature rises, and, at the Davies critical point, 
$\beta_{\rm eff}=\beta_H$,
but as soon as black hole crosses over the Davies point into the region of
$c_Q>0$, the temperature turns lower.

\begin{figure}[t] 
   \centering
   \includegraphics[width=4in]{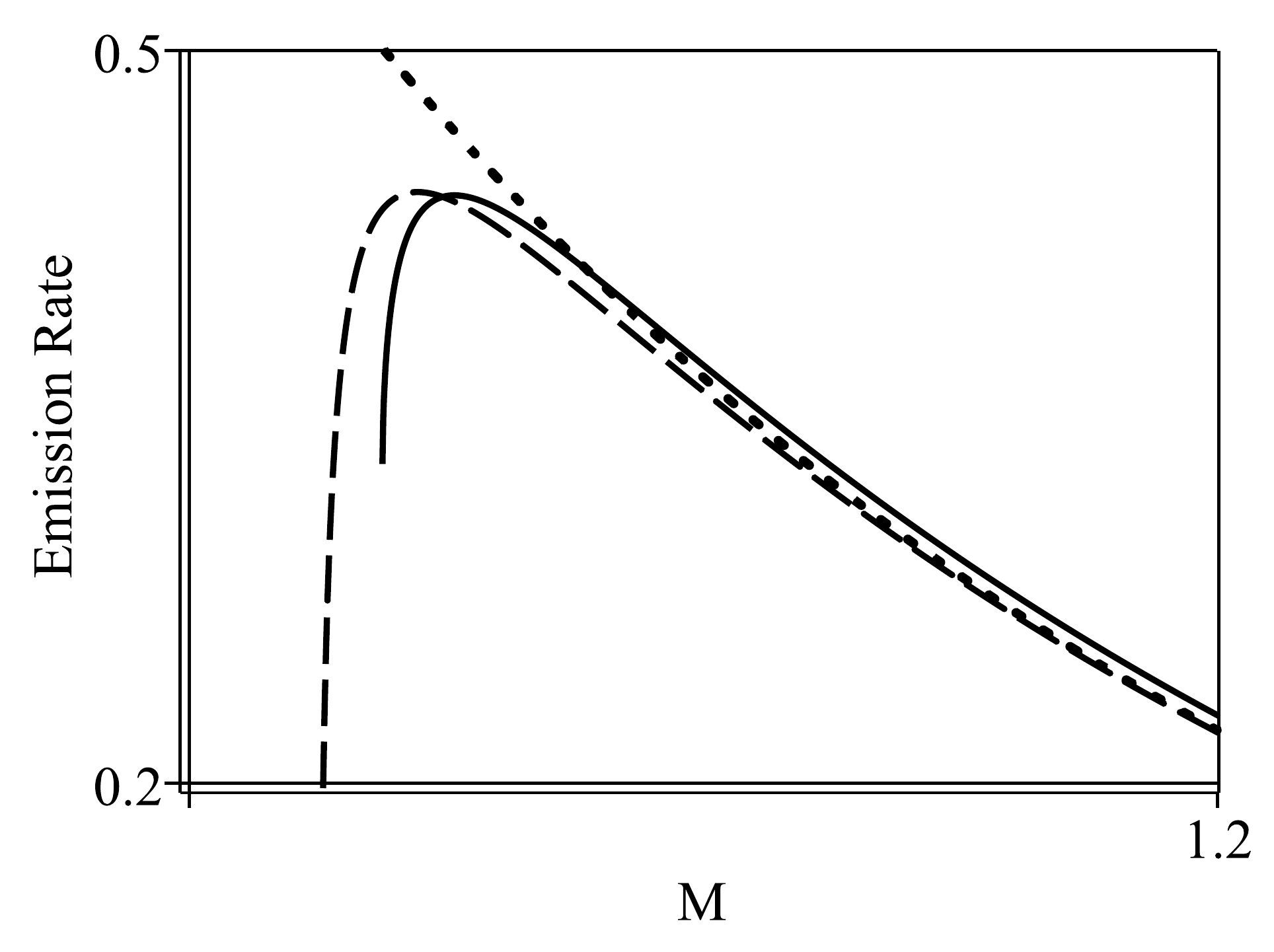} 
   \parbox{0.8\textwidth}{
   \caption{Plots of (normalized) emission rate vs. mass for fixed $Q$. 
   They are exaggerated to visualize
   the difference clearly. The solid line is for RN black hole with 
   nonthermal contribution, while the dashed line is just for thermal 
   part of RN black hole. The dotted line is the thermal part
   of Schwarzschild black hole for comparison.
   As $Q$ increases, the emission rate decreases and 
   the peak location moves to the lower-right direction.}
   \label{fig:errn}}
\end{figure}

This also means the following: Ahead of the Davies point 
(from the right side in fig.\ref{fig:errn}), $c_Q < 0$ and
the (emission rate of) radiation is enhanced by nonthermal radiation 
on top of the thermal radiation. 
At the Davies point, the specific heat is singular and 
the radiation is purely thermal.
But as the black hole crosses the Davies point, $c_Q>0$ and
the radiation is reduced by the nonthermal part, lowering the temperature.
Even though the assumption for this tunneling, $(M-\omega) > Q$,
excludes the extremal case 
(the solid line in fig.\ref{fig:errn} starts off away from the bottom),
but the final formula still seems to make sense in the extremal limit,
where the emission rate does become zero as we expect. So, we can safely
extrapolate this to the extremal limit with caution.
Note that the thermal radiation peaks at the Davies point and the total emission rate peaks at $M=M_c+\omega$. 
Since $\omega$ is small enough compared to $M$, 
it is safe to say that the total radiation peaks at the Davies point.

Compared with the Schwarzschild case, in which there is no such a turning 
critical point, in the charged case, clearly the difference
is caused by the charge $Q$. So the presence of charge is
the cause behind such a phase transition with nonthermal radiation.
Since angular momentum behaves similarly to charge for black holes, 
we can anticipate this structure will persist when $J\neq 0$. 

For the purpose of generalization, let's derive eq.(\ref{eq:beff}) 
more formally. This can be done as follows:
First, note that in terms of Wick rotation $t_{\rm E} = it$
\beq
\label{e:bmbe}
\beta_{\rm M} 
\equiv 2\int_{r_{\rm in}}^{r_{\rm out}}{dr\over \left({dr\over dt}\right)}
= 2\int_{r_{\rm in}}^{r_{\rm out}}{dr\over \left({dr\over -i dt_{\rm E}}\right)} 
=-i\int_0^\beta dt_{\rm E}
= -i\beta,
\eeq
where $\beta$ is the inverse Hawking temperature, which can be shown 
by directly integrating along a geodesic that crosses the horizon.
(See the Appendix \ref{ap:a} for the proof.) 
The factor 2 is present because one needs to cross the horizon
back and forth to complete the period for the Euclidean time to be a
temperature.
Using this, we can identify
\beq
 {\rm Im} \int_M^{M-\omega} \half\beta_{\rm M}(H) dH 
=\half\int_0^\omega \beta(M-\omega')\, d\omega'.
\eeq
Then, expanding in terms of $\omega'$, from eqs.(\ref{e:n1})(\ref{e:n2})
we can obtain
\begin{align}
\beta_{\rm eff}(\omega)\omega
&=\int_0^\omega \beta(M-\omega')\, d\omega'\nonumber\\
&=\int_0^\omega d\omega'
\left(\beta(M) + w'{\pa\beta\over\pa\omega'}\Big|_{\omega'=0}
+ \CO(\omega'^2)\right) \nonumber\\
&= \omega\beta(M) +\half\omega^2{\pa\beta\over\pa\omega}\Big|_{\omega=0}
+ \CO(\omega^3) \nonumber\\
&=\omega\beta(M) + \half{\omega^2(\beta(M))^2\over M c_{Q}}+ \CO(\omega^3),
\end{align}
where $\beta(M) = \beta_H$.
Note that in the above derivation we have not specified what kind of black 
hole we use. So, we have a general theorem that for any black hole
(including the case with $J\neq 0$, i.e. the KN black hole) 
which emits energy $\omega\ll M$, eq.(\ref{eq:beff}) holds true.

The Helmholtz free energy cannot be used to check the stability of the system
with respect to the radiation because it is constant for fixed Hawking 
temperature for which both $Q$ and $M$ need to be fixed. However, we can
take an analogy of the idea of effective temperature and check the stability
with respect to the radiation.
For this purpose, let us define an effective Helmholtz free energy 
after the emission such that
\beq
\label{e:12}
F_{\rm eff} := M_{\rm eff} - T_{\rm eff} S_{\rm eff}.
\eeq
This can be used to compare the Helmholtz free energies before and after
emission of radiation at fixed temperature.
Then using the values after emission
\beq
F_{\rm eff} = M +\Delta M -{1\over\beta_{\rm eff}}(S+\Delta S)
=F +\Delta F,
\eeq
where
\begin{subequations}
\begin{align}
\Delta M &\equiv -\omega <0,\\
\Delta S &\equiv -\beta_{\rm eff}\omega <0,\\
\Delta F &\equiv {1\over 2} {S\over c_{Q}}{\omega\over M}+ \CO(\omega^2)
\end{align}
\end{subequations}
and $F$ is the usual Helmholtz free energy,
hence $\Delta F$ measures the difference at the same Hawking temperature.
In fact, 
\beq
F_{\rm eff} = M-T_{\rm eff}S.
\eeq
Note that the change of the Helmholtz free energy is given in terms of 
the specific heat in the leading order of $\omega/M$.
In particular, $\Delta F=0$ at the critical point.
$\Delta F < 0$ for $c_{Q}<0$, which can be interpreted as that the black hole 
with nonthermal radiation is more stable so that black hole will keep emitting 
nonthermal radiation until it gets evaporated.
However, for $c_Q > 0$, $\Delta F > 0$, which means the black hole before
the nonthermal radiation is more stable, hence the nonthermal radiation 
will be suppressed, or there has to be another mechanism for more nonthermal
emission. This will be addressed in section 5 again.

\section{Davies Critical Point from Tunneling in Kerr-Newman Black Hole}

Next, we shall consider a more general case with $J\neq 0$.
At the same time we shall allow charged 
particles\cite{Zhang:2005uh}\cite{Kerner:2008qv}
(also see \cite{Gibbons:1975kk} for an earlier attempt) 
as well as angular momenta to be emitted.
The thermodynamic relation of a KN blackhole is given by
\beq
\label{e:pt1}
dM = T_H dS + V_H dQ +\Omega_H dJ,
\eeq
where
\begin{subequations}
\begin{align}
T_H ={1\over\beta_H}&={r_+-r_-\over A_K},\\
S &=\frac{1}{4}A_K,\\
V_H &\equiv{4\pi r_+ Q\over A_K},\\
\Omega_H &\equiv{4\pi a\over A_K},
\end{align}
\end{subequations}
where $a=J/M$, $A_K(M,J,Q) = 4\pi(r_+^2 +a^2)$ 
and $r_\pm \equiv M\pm\sqrt{M^2 -Q^2 -a^2}$. 

Then, as shown in Appendix \ref{ap:b}, 
for emission of energy $\omega$, charge $q$, and angular momentum $j$,
we have
\begin{align}
\label{e:19}
-2{\rm Im} I = &\Delta S \nonumber\\
=&-\beta_H\left(\omega - q V_H -j\Omega_H\right)\nonumber\\
&-\half{\omega^2\over M}{\beta_H^2\over c_Q} 
-\half q^2\left({\beta_H\over Q\kappa_Q}
-{\beta_H^2\over \alpha_Q Q} V_H \right)\nonumber\\ 
&-\half j^2\left({\beta_H\over J\kappa_J}
-{\beta_H^2\over \alpha_J J} \Omega_H \right)\nonumber\\ 
&-\omega {q\over Q}{\beta_H^2\over\alpha_Q}
-qj\left({\beta_H^2\over S}\Omega_H V_H
-{\beta_H^2\over \alpha_Q Q}\Omega_H\right)\nonumber\\
&-\omega {j\over J}{\beta_H^2\over\alpha_J} +{\rm h.o.},
\end{align}
where the following definitions of 
thermodynamic quantities for black holes are used:
\begin{subequations}
\begin{align}
&\alpha_Q\equiv {1\over Q}{\pa Q\over\pa T}:
\mbox{charge expansion coefficient},\\
&\alpha_J\equiv {1\over J}{\pa J\over\pa T}:
\mbox{angular momentum expansion coefficient},\\
&\kappa_Q\equiv {1\over Q}\left({\pa Q\over\pa V_H}\right)_T:
\mbox{charge ``compressibility''},\\
&\kappa_J\equiv {1\over J}\left({\pa J\over\pa \Omega_H}\right)_T:
\mbox{angular momentum ``compressibility''}.
\end{align}
\end{subequations}

Note that
\begin{subequations}
\begin{align}
{\beta_H\over Q\kappa_Q}-{\beta_H^2\over \alpha_Q Q} V_H 
&=2\pi\left(1+{1\over B^{1/2}}+{Q^2\over M^2}{1\over B^{3/2}}\right)
>0,\\
{\beta_H\over J\kappa_J}-{\beta_H^2\over \alpha_J J} \Omega_H
&=2\pi{M^2-Q^2\over M^4 B^{3/2}}>0,\\
{\beta_H^2\over \alpha_Q Q}
&=-2\pi{Q(Q^2+a^2)\over M^3 B^{3/2}}
\begin{cases}
<0\ {\rm if}\ Q>0,\\
 >0\ {\rm if}\ Q<0,\\
\end{cases}\\
{\beta_H^2\over S}\Omega_H V_H
-{\beta_H^2\over \alpha_Q Q}\Omega_H
&=2\pi{Qa\over M^3 B^{3/2}}
\begin{cases}
>0\ {\rm if}\ Q>0,\\
 <0\ {\rm if}\ Q<0,\\
\end{cases}\\
{\beta_H^2\over \alpha_J J}
&=-2\pi{a(2M^2-Q^2)\over M^4 B^{3/2}}<0,
\end{align}
\end{subequations}
where $B\equiv 1-(Q^2+a^2)/M^2$ and $J>0$ is assumed without loss of
generality.
This shows that the only term which changes sign for nontrivial values 
of $Q$ and $J$ is the specific heat term. In this case the sign change occurs 
at
\beq
{Q^2\over M^2} = {1\over 4}\left(3 - 6{a^2\over M^2} -{a^4\over M^4}\right),
\eeq
which identifies the Davies critical points\cite{Davies:1978}. 
As pointed out by Davies, this critical phenomenon takes
place even for $Q=0$ at $a^2/M^2 =J^2/M^4= 2\sqrt{3}-3$.
In between, as $a^2/M^2$ varies from zero to $2\sqrt{3}-3$,
$Q^2/M^2$ varies from $3/4$ to zero.

One can also check the signs of other thermodynamic quantities and find
\begin{subequations}
\begin{align}
\alpha_Q &< 0,\\
\alpha_J &<0,\\
\kappa_Q &>0\ {\rm if}\ Q>0,\ <0\ {\rm if}\ Q<0,\\
\kappa_J &>0.
\end{align}
\end{subequations}
In the extremal limit, they all approach to zero, including the specific heat,
and, as $M\to\infty$, they all diverge. So, the existence of the Davies
critical point is quite unique for the specific heat. 

\begin{figure}[t] 
   \centering
   \includegraphics[width=4in]{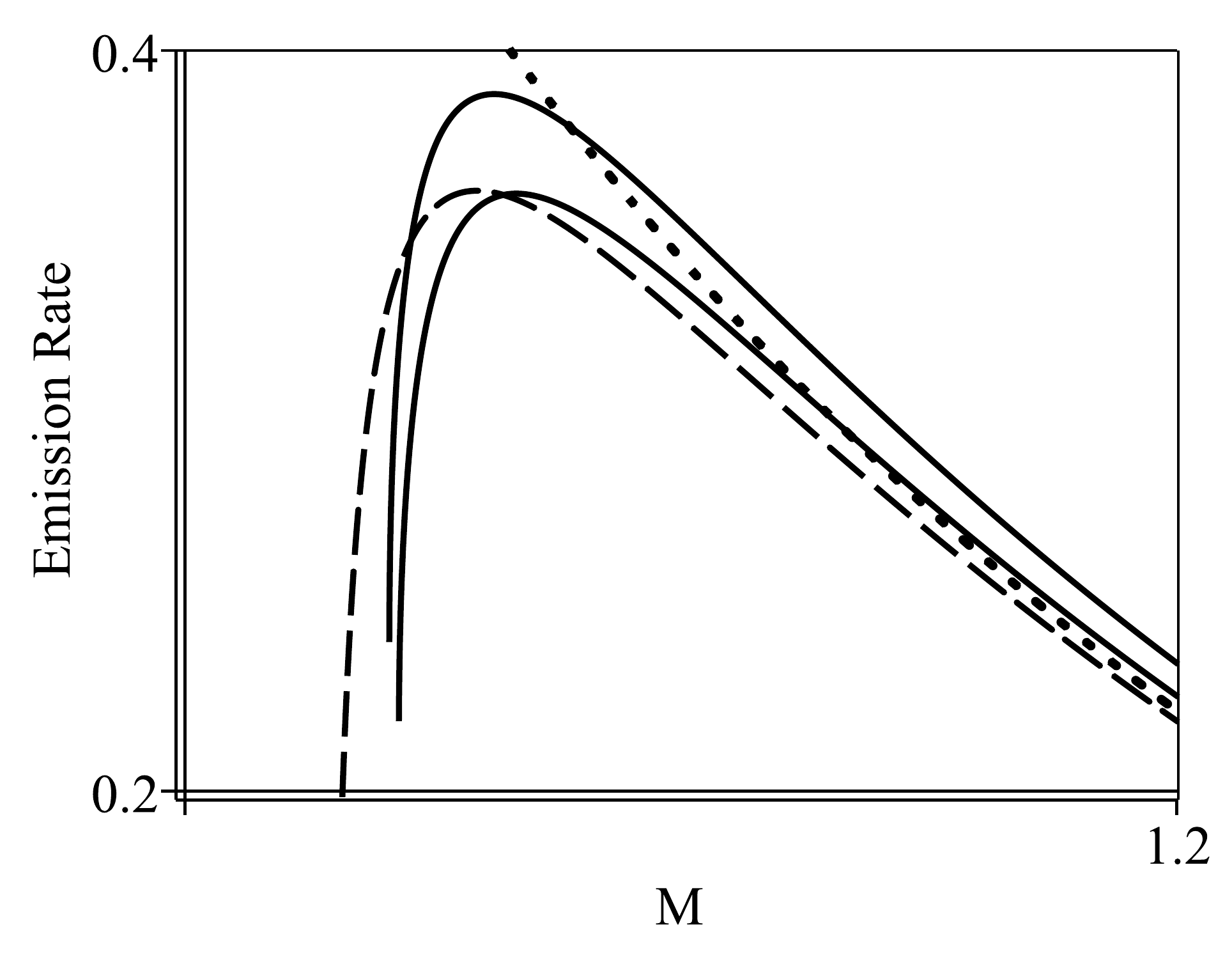} 
   \parbox{0.8\textwidth}{
   \caption{Plots of (normalized) emission rate vs. mass for fixed $Q$ and $J$.
   Again, they are exaggerated to visualize the difference clearly. 
   The lower solid line is for KN black hole with 
   nonthermal contribution for $q=0=j$, 
   and the upper solid line is for KN black hole with 
   nonthermal contribution for $q\neq 0\neq j$, $qQ>0$,
   while the dashed line is just for thermal 
   part of KN black hole. The dotted line is the thermal part
   of Schwarzschild black hole for comparison.
   As $Q$ or $J$ increases, the emission rate decreases and 
   the peak location moves to the lower-right direction.}
   \label{fig:erknqj}}
\end{figure}

Even though $\omega,q,j$, are introduced independently, they
cannot be arbitrary. This is because physically
charges or angular momenta cannot be emitted without changing energy.
For example, it is necessary that 
\beq
\label{e:oq}
(\omega-qV_H-j\Omega_H)>0
\eeq
for the emission rate to make sense and that $\Delta S <0$. 
Otherwise, the emission rate diverges as temperature goes to zero. 
This provides us interesting bounds. 

Let's consider $j=0$ case, first. Then 
\beq
{\omega\over |q|}> |V_H| 
= {|Q|\over M}{1+\sqrt{B}\over (1+\sqrt{B})^2 +a^2/M^2},
\eeq
for $qQ>0$. $|V_H|<1$ and approaches zero as $M\to\infty$, hence
this does not necessarily imply $\omega > q$. Nevertheless, this bound is quite
significant, compared with $m_e/e\sim 10^{-21}$ or $m_p/e\sim 10^{-18}$.
This means that charge has to be emitted with a sufficiently large energy.
Not surprisingly, the larger $Q/M$ ratio is, 
the more energetically charges will be emitted.
This also has a rather interesting alternative interpretation. 
Suppose $Q/M$ violates this bound for some charge emission,
then the bad behavior of emission rate can be interpreted as black holes
inability to sustain charge-to-mass ratio $Q/M$. There are two ways of lowering
$Q/M$: either emit charges or absorb more mass to increase $M$.
What if not much matter is around the black hole to absorb? 
The only resolution will be to emit charges. In other words,
this is a bound for a black hole how much charge it can sustain.
Based on the classical argument, we know $Q/M < m/q$ is a limit for 
a black hole to absorb stationary charged particle. 
Here we have another bound beyond which a black hole cannot emit charges
normally, but must get rid of them drastically. Note that, unlike
the classical bound, the bound in the tunneling case is given in terms of 
emitting energy, not rest mass. 
If $qQ<0$, eq.(\ref{e:oq}) is always satisfied. This means that 
opposite charges can also be emitted and $Q$ can actually increase,
although the emission rate is lower than that of $qQ>0$.

A similar bound can be obtained for the emission of the angular momenta.
For $q=0$, since $jJ>0$,
\beq
{\omega\over j}> \Omega_H
= {J\over M^3}{1\over (1+\sqrt{B})^2 +J^2/M^4}.
\eeq

To look into the next order contribution, we first need to
impose eq.(\ref{e:oq}).
Now the emission rate actually depends on the signs of $Q$ and $q$.
The $q^2$ and $j^2$ terms always enhance the emission rate,
but the contribution of $\omega^2$ term depends on the sign of specific heat
and vanishes at the Davies point, where the emission rate peaks.
If $qQ>0$, the last two terms with $q$
in the quadratic expansion enhance the emission rate.
However, if $qQ<0$, they suppress the emission rate.
The term with $j$ always enhances the emission rate since we assume $J>0$
and $j>0$.

In the RN case, we checked the stability of the black hole with respect 
to the radiation of energy, using the effective Helmholtz free energy. 
In the KN case, however, the effective Helmholtz free energy is not sufficient 
because of charge and angular momentum emissions.
Note that the Gibbs free energy cannot be used to check the stability
of a KN black hole with respect to the radiation either because it is 
constant for fixed $T_H$, $V_H$ and $\Omega_H$, for which all $M,Q, J$
are fixed.

So, we need to consider the effective Gibbs free energy analogously
defined, like the effective Helmholtz free energy, 
in terms of effective quantities as
\beq
\label{e:efg}
G_{\rm eff} = M_{\rm eff} - T_{\rm eff} S_{\rm eff}
-V_{\rm eff}Q_{\rm eff} -\Omega_{\rm eff} J_{\rm eff},
\eeq 
then the difference between the effective Gibbs free energy
and the usual Gibbs free energy at the same Hawking temperature 
reads, as derived in Appendix \ref{ap:c},
\beq
\label{e:29}
\Delta G \equiv G_{\rm eff} -G =\omega G_\omega +q G_q +j G_j,
\eeq
where
\begin{subequations}
\begin{align}
G_\omega &\equiv -{\pi^2\over\beta_H^2}{A\over B^{3/2}}
\left(2M^2\! +Q^2\! +{2M^2\over B^{1/2}}\right),
\\
G_q &\equiv {\pi Q\over\beta_H B^{3/2}}\nonumber\\
&\!\times\!\left(1+B^{3/2}\! -{2\pi\over\beta_H}
{Q^2+2J^2/M^2\over M^3}\!\left(2M^2\! +\! Q^2\! 
+\!{2M^2\over B^{1/2}}\right)\!
\right),
\\
G_j &\equiv {\pi J\over\beta_H B^{3/2}}\!\left(\! {1\over M^2}
\! -\! {2\pi\over\beta_H}
{2M^2\! -\! Q^2\over M^5}\!\left(2M^2\! +Q^2\! +{2M^2\over B^{1/2}}\right)\!
\right),
\end{align}
\end{subequations}
where $B\equiv 1-(Q^2+J^2/M^2)/M^2$ as before and
\beq
A\equiv 2B^{3/2} +2 -6{J^2\over M^4} -3{Q^2\over M^2} +{J^2 Q^2\over M^6}.
\eeq
Note that $A=0$ identifies the Davies critical points.
One can show that $G_j<0$ for $J\neq 0$ always,
however, the signs of $G_\omega$ and $G_q$ depend on $Q,J$ and $M$.

In summary, there are over all three different cases of $\Delta G$ 
depending on the variables, which are $M,Q,J$ and $\omega,q,j$:
$\Delta G <0$ for all $M$, $\Delta G\gtrless 0$ for $M\lessgtr M_0$,
where $M_0$ is another solution to $\Delta G=0$,
or $\Delta G>0$ for some interval of $M$ while $\Delta G <0$ for the rest $M$.
If $J=0$, $M_0=M_c$, but if $J\neq 0$, $M_0$ is different from $M_c$.
The existence of opposite signs of $\Delta G$ indicates 
there are two different phases of KN black holes.
The details are as follows. 

If $J=0$, this describes the RN black hole with charge emission for
$q\neq 0$ or without charge emission for $q=0$.
In the $q=0$ case, $\Delta G$ correctly reduces to $\Delta F$ so that 
it simply reproduces the result of effective Helmholtz free energy case.
So we shall consider $q\neq 0$ case here.
For $J=0$, $G_q$ can be rewritten as
\beq
G_q = {\pi Q\over \beta_H B^{3/2}}
{2M^2(1+B^{1/2})-Q^2\over(2M^2-Q^2+2M^2 B^{1/2})(1+B^{1/2})} A
\eeq
so that $G_q = 0$ if $A=0$. $G_\omega = 0 = G_q$, hence $\Delta G = 0$
at $M=M_c$, that is, the Davies critical point.
For $J=0$ and fixed $Q$,
$G_\omega \gtrless 0$ for $M\lessgtr M_c$, while
$G_q \lessgtr 0$ for for $M\lessgtr M_c$. 
So, if $\omega/q \geq 1$, $\Delta G \gtrless 0$ for $M\lessgtr M_c$.
In the case $\omega/q<1$, there is a value $q'$ such that,
if $\omega/q < \omega/q'<1$, $J=0$, $\Delta G> 0$ for 
$M_c > M> M_0$, while, if $\omega/q' < \omega/q<1$, $J=0$, $\Delta G> 0$ 
for $M_0 > M> M_c$, where $M_0$ is again another solution to $\Delta G=0$
and if $\omega/q=\omega/q' <1$, $M_0=M_c$, hence $\Delta G\leq 0$.
So, $M_0$ depends on the ratio $\omega/q$, which is another indication
that $\omega$ and $q$ are not totally independent.

If $J\neq 0$, $G_\omega$ still vanishes at $M_c$, but
$G_q$ vanishes at some other value $M'$ different from $M_c$. 
For $j=0$ and $\omega/q>\omega/q'$, $\Delta G\gtrless 0$
for $M\lessgtr M_0$ and $\Delta G =0$ for $M=M_0$.
For $j=0$ and $\omega/q<\omega/q' <1$, 
$\Delta G > 0$ for $M_0>M> M_{\rm min}$, where $M_{\rm min}$ is
the bound given by eq.(\ref{e:oq}) and $q'$ in this case 
is given by $\Delta G=0$ for $M=M_{\rm min}=M_{\rm ex}$, where 
$M_{\rm ex}$ is the value of the extremal black hole such that
$M^2_{\rm ex}\equiv\half(Q^2 +\sqrt{Q^2 +4 J^2})$.

If $J\neq 0$ and $j\neq 0$, if $\omega/q >\omega/q'$, $\Delta G <0$ for all 
$M$, while if $\omega/q=\omega/q' <1$ and $j<j'$, $\Delta G>0$
for $M_0>M>M_{\rm min}$ and $\Delta G<0$ for $M>M_0$ or $M>M_{\rm min}$. 
However, as $j$ gets larger, the gap between
$M_0$ and $M_{\rm min}$ shrinks, at that, if $j>j'$ for some $j'$, 
$\Delta G<0$ for all $M$ again.

If $J\neq 0$ and $q=0$, for $\omega/j >\omega/j'$, $\Delta G\gtrless 0$
for $M\lessgtr M_0$, while for $\omega/j <\omega/j'$, $\Delta G<0$ for all
$M$, where $j'$ is determined by $\Delta G=0$ for $M_0=M_{\rm ex}$.
The $Q=0$ and $J\neq 0$ case has a similar behavior, too. 

\section{Discussion}

We have shown the physical implication of the Davies critical phenomenon
from the tunneling point of view. The sign of specific heat is related
to the sign of the nonthermal contribution to the black hole radiation.
We have also shown that the emission rate based on the tunneling peaks 
at (or near, if charge or angular momentum emissions are allowed) 
the Davies critical point. The argument based on the
effective free energies show
there are two phases, indicating whether the emission is spontaneous or not.

Negative specific heat is common in collapsing self-gravitating 
systems which are isolated\cite{LyndenBell:1998fr}, 
so the real question to address in the black hole context is
what happens to the black hole when the specific heat is positive.
A black hole can never be truly a stable thermodynamic system
because either matter always falls in and/or radiations always come out.
The equilibrium between falling matter and outward emission does not 
save the situation because of the case without matter available to fall
and black hole solution does not distinguish it.
If isolated, a black hole cannot evolve adiabatically without
violating the charge conservation or angular momentum conservation.
In case radiations come out at the same time matter falls,
we cannot use tunneling argument to test the adiabatic evolution 
because the emission rate 
remains constant for any temperature, hence does not make sense.
One could argue based on thermodynamic relations, 
using $0=dS=dM-V_H dQ -\Omega_H dJ$, to check if there is a solution
with a reasonable boundary condition.
However, in the limit $Q$ and $J$ vanish, $M$ has to be constant.
Since $M$ cannot be zero because, otherwise, it violates the unitarity,
we end up with a Schwarzschild black hole whose specific heat is always 
negative.
So the positive sign of the specific heat does not really imply its stability,
unlike normal thermodynamic systems. 
Even in the extremal limit, it
is not perfectly clear if the black hole is stable. 
(For example, see \cite{Kab}.) $T_H=0$, but we still
have high charge-to-mass or angular-momentum-to-mass ratio so that it is not
clear how it can sustain the zero temperature. We expect there should be no
radiation, but the tunneling argument based on the WKB method fails
in the extremal limit so that we cannot confirm, although it appears to be
consistent.
It will be interesting if one can check whether a black hole can really evolve
into the extremal limit, or some kind of phase transition takes place 
at the extremal limit and it actually fails to be a black hole.

From the tunneling point of view, the positive specific heat indicates
that the radiation is suppressed by the nonthermal contribution. The
effective Helmholtz free energy leads to $\Delta F>0$, indicating this
suppression does not take place spontaneously. We suspect that this signals
there should be another phenomenon taking place presumably right outside
of the horizon since it may not be an intrinsic black hole phenomenon, 
which will end up increasing the black hole
radiation toward the thermal level direction,
 like that of Damour-Ruffini\cite{Damour:1974qv}
and Blandford-Znajek\cite{Blandford:1977ds}.
Therefore, we also suspect that there might be a connection between 
Parikh-Wilczek's nonthermal radiation\cite{Parikh:1999mf} and 
that of Damour-Ruffini\cite{Damour:1974qv}
and Blandford-Znajek\cite{Blandford:1977ds}.
The effort to relate them is in progress and will be reported elsewhere.

\acknowledgments{
The author thanks L. Alvarez-Gaum\'e and S. Deser for helpful communications,
and G. Siopsis for helpful discussions.
}

\appendix

\section{Yet Another Derivation of Hawking Temperature}\label{ap:a}

The black hole temperature can be obtained by performing the Wick rotation of
the metric so that the period of the
Euclidean time along a circle can be identified as the inverse temperature. 
Consider a generic metric with the Euclidean signature
\beq
\label{emetuv}
ds^2 = U(r)V(r)dt^2 + {1\over V(r)}dr^2 + r^2 d\Omega^2.
\eeq
A godly traveler moving along a geodesic should be back to 
the original location after this time period, 
crossing the horizon back and forth.
So we can compute the traveling period $\beta$ as
\beq
\label{eq:beta}
\beta = \int_0^\beta dt 
=\oint {dt\over dr}dr =2\int_P {dt\over dr}dr
=2\int_P {dr\over \dot{r}},
\eeq
where $P$ denotes a geodesic path crossing the horizon and we have 
eq.(\ref{e:bmbe}) with $t\equiv t_{\rm E}$. 
Note that for $ds^2=0$ and $\theta=0=\phi$
\beq
\label{e:nul}
{dt\over dr}=\sqrt{-{g_{rr}\over g_{tt}}}.
\eeq
So we might be tempted just to use this, but in the Euclidean signature
this is imaginary so that we need to be a little bit more cautious
and justify it.

Without loss of generality and for convenience, we choose $\theta=0=\phi$,
then nontrivial geodesic equations are
\begin{subequations}
\begin{align}
\label{eq:geoda}
0 &=\ddot{t}+\left(\ln(UV)\right)'\dot{t}\dot{r}, \\
\label{eq:geodb}
0 &=\ddot{r} -\half V(UV)'\dot{t}^2 -\half {V'\over V}\dot{r}^2,
\end{align}
\end{subequations}
where the dot denotes the derivative with respect to the affine 
parameter $\lambda$ and the prime denotes the derivative with respect to $r$.
Eq.(\ref{eq:geoda}) can be integrated to
\beq
\label{eq:ta}
UV\dot{t}=c_1,
\eeq
where $c_1$ is an integration constant. 
Using this, eq.(\ref{eq:geodb}) can be written as
\beq
0 =\ddot{r} -\half V{(UV)'\over (UV)^2}c_1^2 -\half {V'\over V}\dot{r}^2.
\eeq
Multiplying $2\dot{r}/V$, we obtain 
\beq
0={d\over d\lambda}\left({\dot{r}^2\over V}\right)+c_1^2{d\over d\lambda}
\left({1\over UV}\right),
\eeq
hence
\beq
\label{eq:tb}
\dot{r}^2 +{c_1^2\over U}=c_2 V,
\eeq
where $c_2$ is another integration constant.
Using eq.(\ref{eq:ta}) and eq.(\ref{eq:tb}), we can perform the integration 
eq.(\ref{eq:beta}):
\beq
\beta=2\int_P {dt\over dr}dr 
= 2\int_P dr{c_1\over UV}{1\over\sqrt{c_2V -c_1^2/U}}.
\eeq
Note that we can choose $c_2=0$ for our geodesic path 
such that eq.(\ref{e:nul}) is more or less justified.
This integral is multi-valued, so
we should choose the smallest nonvanishing value as the period. 
Then the integration simply
becomes (assuming the pole is a simple pole, which is usually the case),
using the Cauchy theorem after analytic continuation with a suitable path,
\begin{align}
\beta&={2\over i}\int_P dr{1\over \sqrt{U}V}\nonumber\\
&={2\over i}2\pi i\,{\mathop{}^{\textstyle\rm Res}_{r=r_H}}{1\over \sqrt{U}V}
\nonumber\\
&=4\pi\lim_{r\to r_H}\left[{d\over dr}\left({\sqrt{U}V}\right)\right]^{-1}.
\end{align}
The black hole temperature is, then, given by
\beq
\label{ehtuv}
T_H={1\over \beta}
={1\over 4\pi}\lim_{r\to r_H}{d\over dr}\left({\sqrt{U}V}\right)
={1\over 4\pi}\lim_{r\to r_H}{d\over dr}\sqrt{g_{tt}\over g_{rr}}.
\eeq
For $U=1$, we have
\beq
T_H={1\over \beta} 
={1\over 4\pi} \lim_{r\to r_H}{d\over dr}g_{tt}.
\eeq

For KN black hole, set $\theta=0$, which in turn sets
$g_{t\phi}=0$ (also set $g_{\phi\phi}=0$ at the end), 
then the metric satisfies the same geodesic equations
eqs.(\ref{eq:geoda},\ref{eq:geodb}).
So we can still apply eq.(\ref{ehtuv}) to obtain the Hawking temperature 
for KN black hole as
\beq
T_H={r_+ - r_-\over 4\pi(r_+^2 + a^2)}.
\eeq

\section{Derivation of Eq.(\ref{e:19}) for KN black hole}\label{ap:b}

In the WKB approximation, using the Hamilton-Jacobi theory, 
the emission of energy $\omega$, charge $q$, and angular momentum $j$,
is given by\cite{Zhang:2005uh}\cite{Kerner:2008qv}
\beq
\label{eq:s}
{\rm Im} I
={\rm Im}\int\left(H-p_A\dot{A} -p_\phi\dot{\phi}\right)dt,
\eeq
where the generalized coordinates we use are $(r, A, \phi)$. 
Using eq.(\ref{e:pt1}) and the fact that, as far as these generalized coordinates are concerned, $S, Q, J$ are corresponding independent variables, 
we can obtain the following identities these generalized 
coordinates should satisfy:
\begin{subequations}
\begin{align}
\label{e:hja}
\dot{r} &= {\pa H\over\pa p_r}={1\over\beta_H}{\pa S\over \pa p_r},\\
\label{e:hjb}
\dot{A} &={\pa H\over\pa p_A}=V_H{\pa Q\over \pa p_A},\\
\label{e:hjc}
\dot{\phi} &={\pa H\over\pa p_\phi}=\Omega_H{\pa J\over \pa p_\phi}.
\end{align}
\end{subequations}
Substituting the integrations $H=\int dM$, $p_A=\int dp_A$, 
and $p_\phi=\int dp_\phi$, we obtain 
\beq
{\rm Im}I={\rm Im}\iint {dr\over \dot{r}}\left(dM-V_H dQ-\Omega_H dJ\right).
\eeq
In fact, using eq.(\ref{e:hja}), this is actually equivalent to
\begin{align}
\label{eq:pr}
{\rm Im} I &={\rm Im}\int p_r \dot{r} dt \nonumber\\
&={\rm Im}\int p_r {\pa H\over\pa p_r} dt \nonumber\\
&={\rm Im}\iint dp_r{1\over\beta_H}{\pa S\over \pa p_r}dt \nonumber\\
&={\rm Im}\iint{dr\over\dot{r}}{1\over\beta_H}dS.
\end{align}
From eq.(\ref{eq:s}) and eq.(\ref{eq:pr}), we can consistently identify $H$ as 
the Hamilton's characteristic function in this case.

Finally, using eq.(\ref{e:bmbe}), we can show that
\beq
-2{\rm Im} I = \Delta S \equiv\int dS=S(M-\omega,Q-q,J-j)-S(M,Q,J).
\eeq
We can assume $J>0$ without loss of generality and that $jJ>0$.
However, the formula is valid for any signs of $Q$ and $q$, as long as
$\Delta S <0$, hence we are not going to make any assumption 
on the signs of $Q$ and $q$ at this moment.
Let's expand $\Delta S$ in terms of $\omega,q,j$, such that
\begin{align}
\label{e:8}
\Delta S = &-\omega{\pa S\over \pa M} -q{\pa S\over \pa Q}
-j{\pa S\over \pa J}\nonumber\\
&+\half\omega^2{\pa^2 S\over \pa M^2}
+\half q^2{\pa^2 S\over \pa Q^2}
+\half j^2{\pa^2 S\over \pa J^2}\nonumber\\
&+\omega q{\pa^2 S\over \pa M\pa Q}
+ q j{\pa^2 S\over \pa Q\pa J}
+j\omega {\pa^2 S\over \pa J\pa M}
+{\rm h.o.}
\end{align}
Now define the following thermodynamic quantities for black holes:
\begin{subequations}
\begin{align}
&\alpha_Q\equiv {1\over Q}{\pa Q\over\pa T}:
\mbox{charge expansion coefficient},\\
&\alpha_J\equiv {1\over J}{\pa J\over\pa T}:
\mbox{angular momentum expansion coefficient},\\
&\kappa_Q\equiv {1\over Q}\left({\pa Q\over\pa V_H}\right)_T:
\mbox{charge ``compressibility''},\\
&\kappa_J\equiv {1\over J}\left({\pa J\over\pa \Omega_H}\right)_T:
\mbox{angular momentum ``compressibility''}.
\end{align}
\end{subequations}
Then, from eq.(\ref{e:pt1}), 
we can identify the following thermodynamic relations:
\begin{subequations}
\begin{align}
{\pa S\over\pa M} &=\beta_H,\\
{\pa S\over\pa Q} &= -\beta_H V_H,\\
{\pa S\over\pa J} &= -\beta_H \Omega_H,\\
{\pa^2 S\over\pa M^2} &= -{\beta_H^2\over M c_Q},\\
{\pa^2 S\over\pa Q^2} &=-{\pa\over\pa Q}(\beta_H V_H)
={\beta_H^2\over \alpha_Q Q} V_H -{\beta_H\over Q\kappa_Q},\\
{\pa^2 S\over\pa J^2} &=-{\pa\over\pa J}(\beta_H \Omega_H)
={\beta_H^2\over \alpha_J J} \Omega_H -{\beta_H\over J\kappa_J},\\
{\pa^2 S\over\pa M\pa Q} 
&={\pa\beta_H\over\pa Q} = -{\beta_H^2\over \alpha_Q Q},\\
{\pa^2 S\over\pa Q\pa J} 
&=-{\pa\over\pa Q}(\beta_H\Omega_H) 
= {\beta_H^2\over \alpha_Q Q}\Omega_H
-{\beta_H^2\over S}\Omega_H V_H,\\
{\pa^2 S\over\pa J\pa M} 
&={\pa\beta_H\over\pa J} = -{\beta_H^2\over \alpha_J J}.
\end{align}
\end{subequations}
Thus we have 
\begin{align}
-2{\rm Im} I = &\Delta S \nonumber\\
=&-\beta_H\left(\omega - q V_H -j\Omega_H\right)\nonumber\\
&-\half{\omega^2\over M}{\beta_H^2\over c_Q} 
-\half q^2\left({\beta_H\over Q\kappa_Q}
-{\beta_H^2\over \alpha_Q Q} V_H \right) \nonumber\\
&-\half j^2\left({\beta_H\over J\kappa_J}
-{\beta_H^2\over \alpha_J J} \Omega_H \right) \nonumber\\ 
&-\omega {q\over Q}{\beta_H^2\over\alpha_Q}
-qj\left({\beta_H^2\over S}\Omega_H V_H
-{\beta_H^2\over \alpha_Q Q}\Omega_H\right) \nonumber\\
&-\omega {j\over J}{\beta_H^2\over\alpha_J} +{\rm h.o.},
\end{align}
which is eq.(\ref{e:19}).

\section{Derivation of Eq.(\ref{e:29}) for KN black hole}\label{ap:c}

The effective Gibbs free energy is defined by
\beq
\label{e:gb}
G_{\rm eff} = M_{\rm eff} - T_{\rm eff} S_{\rm eff}
-V_{\rm eff}Q_{\rm eff} -\Omega_{\rm eff} J_{\rm eff}.
\eeq 
$M_{\rm eff}$ and $S_{\rm eff}$ can be identified as before in eq.(\ref{e:12}),
and the rest can be identified as following.
The natural identification of $\Delta S$ in this case should be
\beq
\Delta S = -\beta_{\rm eff}\left(\omega -qV_{\rm eff}-j\Omega_{\rm eff}\right),
\eeq 
then from eq.(\ref{e:8}) we can read off the effective quantities as
\begin{subequations}
\begin{align}
\beta_{\rm eff} &={\pa S\over \pa M} -\half\omega{\pa^2 S\over \pa M^2}
-\left(q{\pa^2 S\over \pa M\pa Q}+j{\pa^2 S\over \pa J\pa M} \right),\\
\beta_{\rm eff}V_{\rm eff} &=-{\pa S\over \pa Q}
+\half q{\pa^2 S\over \pa Q^2}
+\half j{\pa^2 S\over \pa Q\pa J},\\
\beta_{\rm eff}\Omega_{\rm eff} &=-{\pa S\over \pa J}
+\half j{\pa^2 S\over \pa J^2}
+\half q{\pa^2 S\over \pa Q\pa J}.
\end{align}
\end{subequations}
These identifications consistently truncate the effective Gibbs free energy
to the effective Helmholtz free
energy. And the definition of the effective Gibbs free energy eq.(\ref{e:gb})
also actually reduces to
\beq
G_{\rm eff} = M - T_{\rm eff} S -V_{\rm eff}Q -\Omega_{\rm eff} J.
\eeq
Now we can compute the difference between the effective Gibbs free energy
and the usual Gibbs free energy at the same Hawking temperature as
\beq
\Delta G \equiv G_{\rm eff} -G
={S\over \beta_H^2}\Delta\beta -Q\Delta V -J\Delta\Omega,
\eeq
where
\begin{subequations}
\begin{align}
\Delta\beta 
&= \beta_{\rm eff}-\beta_H \nonumber\\
&= \beta_H^2\left(\half{\omega\over M}{1\over c_Q}
+{q\over Q}{1\over \alpha_Q}+{j\over J}{1\over \alpha_J}\right),\\
\Delta V 
&= V_{\rm eff}-V_H \nonumber\\
&=
-\half {q\over Q}\left({1\over \kappa_Q}
-{\beta_H\over \alpha_Q} V_H \right) 
-\half j\beta_H\Omega_H\left({V_H\over S}-{1\over \alpha_Q Q}\right)\nonumber\\
&\quad -{V_H\over\beta_H}\Delta\beta,\\
\Delta\Omega 
&= \Omega_{\rm eff}-\Omega_H \nonumber\\
&=
-\half {j\over J}\left({1\over \kappa_J}
-{\beta_H\over \alpha_J} \Omega_H \right) 
-\half q\beta_H\Omega_H\left({V_H\over S}-{1\over \alpha_Q Q}\right)\nonumber\\
&\quad -{\Omega_H\over\beta_H}\Delta\beta.
 \end{align}
\end{subequations}

This can be rewritten as
\beq
\Delta G = \omega G_\omega +q G_q +j G_j,
\eeq
where
\begin{widetext}
\begin{subequations}
\begin{align}
G_\omega &\equiv -{\pi^2\over\beta_H^2}{A\over B^{3/2}}
\left(2M^2+Q^2+{2M^2\over B^{1/2}}\right),
\\
G_q &\equiv {\pi Q\over\beta_H B^{3/2}} \left(1+B^{3/2} 
-{2\pi\over\beta_H}
{Q^2+2J^2/M^2\over M^3}\left(2M^2+Q^2+{2M^2\over B^{1/2}}\right)
\right),
\\
G_j &\equiv {\pi J\over\beta_H B^{3/2}}\left({1\over M^2}
-{2\pi\over\beta_H}
{2M^2-Q^2\over M^5}\left(2M^2+Q^2+{2M^2\over B^{1/2}}\right)
\right),
\end{align}
\end{subequations}
\end{widetext}
where
\begin{subequations}
\begin{align}
B&\equiv 1-{Q^2 +J^2/M^2\over M^2},\\
A&\equiv 2B^{3/2} +2 -6{J^2\over M^4} -3{Q^2\over M^2} +{J^2 Q^2\over M^6}.
\end{align}
\end{subequations}


\begin{thebibliography}{100}

\bibitem{Bekenstein:1973ur}
  J.~D.~Bekenstein,
  Phys.\ Rev.\  D {\bf 7}, 2333 (1973).

\bibitem{Davies:1978}
 P.~C.~W.~Davies,
Rep. Prog. Phys. {\bf 41} 1313 (1978);
  Proc.\ Roy.\ Soc.\ Lond.\  A {\bf 353}, 499 (1977).
 
\bibitem{Hut}
P. Hut,
Mon. Not. R. Astr. Soc. {\bf 180} 379 (1977).


\bibitem{Hiscock:1990ex}
  W.~A.~Hiscock and L.~D.~Weems,
  Phys.\ Rev.\  D {\bf 41}, 1142 (1990).

\bibitem{Kaburaki:1993ah}
  O.~Kaburaki, I.~Okamoto and J.~Katz,
  Phys.\ Rev.\  D {\bf 47}, 2234 (1993).

\bibitem{Parikh:1999mf}
  M.~K.~Parikh and F.~Wilczek,
  Phys.\ Rev.\ Lett.\  {\bf 85}, 5042 (2000)
  [arXiv:hep-th/9907001].

\bibitem{Parikh:2004rh}
  M.~K.~Parikh,
  arXiv:hep-th/0402166.

\bibitem{Damour:1974qv}
  T.~Damour and R.~Ruffini,
  Phys.\ Rev.\ Lett.\  {\bf 35}, 463 (1975).

\bibitem{Blandford:1977ds}
  R.~D.~Blandford and R.~L.~Znajek,
  Mon.\ Not.\ Roy.\ Astron.\ Soc.\  {\bf 179}, 433 (1977).

\bibitem{Hawking:1974sw}
  S.~W.~Hawking,
  Commun.\ Math.\ Phys.\  {\bf 43}, 199 (1975)
  [Erratum-ibid.\  {\bf 46}, 206 (1976)].


\bibitem{Zhang:2005uh}
  J.~Zhang and Z.~Zhao,
  Phys.\ Lett.\  B {\bf 638}, 110 (2006)
  [arXiv:gr-qc/0512153].

\bibitem{Kerner:2008qv}
  R.~Kerner and R.~B.~Mann,
  Phys.\ Lett.\  B {\bf 665}, 277 (2008)
  [arXiv:0803.2246 [hep-th]].

\bibitem{Gibbons:1975kk}
  G.~W.~Gibbons,
  Commun.\ Math.\ Phys.\  {\bf 44}, 245 (1975).

\bibitem{LyndenBell:1998fr}
D.~Lynden-Bell,
arXiv:cond-mat/9812172.

\bibitem{Kab}
O. Kaburaki,
Gen. Rel. Grav. {\bf 23} (1996) 843.    

\end{thebibliography}
\end{document}